\documentclass[
 reprint,
 superscriptaddress,
 amsmath,amssymb,
 aps,
 floatfix,
]{revtex4-2}

\usepackage{graphicx}
\usepackage{dcolumn}
\usepackage{bm}
\usepackage{hyperref}

\usepackage{csquotes}
\usepackage{physics}
\usepackage{simpler-wick}
\usepackage{tikz}
\usepackage{accents}
\usepackage{svg}
\usepackage{mathtools}
\usepackage[percent]{overpic}

\newcommand{\singlet}{
\begin{tikzpicture}[baseline=-0.5ex]
\begin{scope}[scale=0.8]
\draw (0,0) ellipse (0.1 and 0.25);
\end{scope}
\end{tikzpicture}
}

\newcommand{\cc}{
\begin{tikzpicture}[baseline=-0.5ex]
\begin{scope}[scale=0.8]
\draw (0,-0.1) circle (0.1);
\draw (0,0.2) circle (0.10);
\end{scope}
\end{tikzpicture}
}

\newcommand{\scup}{
\begin{tikzpicture}[baseline=-0.5ex]
\begin{scope}[scale=0.8]
\draw (0,-0.1) node[above=2] {$\sigma$} circle (0.1);
\end{scope}
\end{tikzpicture}
}

\newcommand{\scdown}{
\begin{tikzpicture}[baseline=-0.5ex]
\begin{scope}[scale=0.8]
\draw (0,+0.25) node[below=2] {$\sigma$} circle (0.1);
\end{scope}
\end{tikzpicture}
}

\newcommand\thickbar[1]{\accentset{\rule{.3em}{.5pt}}{#1}}

\begin{document}

\title{Extended $s$-wave pairing from an emergent Feshbach resonance \\ in bilayer nickelate superconductors}

\author{Pietro Borchia}
\affiliation{ Ludwig-Maximilians-University Munich, Theresienstr. 37, Munich D-80333, Germany
}%
\affiliation{ Technical University of Munich, TUM School of Natural Sciences, Physics Department, 85748 Garching, Germany
}%
\author{Hannah Lange}
\affiliation{ Ludwig-Maximilians-University Munich, Theresienstr. 37, Munich D-80333, Germany
}%
\affiliation{Max-Planck-Institute for Quantum Optics, Hans-Kopfermann-Str.1, Garching D-85748, Germany}
\affiliation{Munich Center for Quantum Science and Technology, Schellingstr. 4, Munich D-80799, Germany}
\author{Fabian Grusdt}
\affiliation{ Ludwig-Maximilians-University Munich, Theresienstr. 37, Munich D-80333, Germany
}%
\affiliation{Munich Center for Quantum Science and Technology, Schellingstr. 4, Munich D-80799, Germany}

\date{\today}

\begin{abstract}
Since the discovery of unconventional superconductivity in cuprates, unraveling the pairing mechanism of charge carriers in doped antiferromagnets has been a long-standing challenge. Motivated by the discovery of high-T$_c$ superconductivity in nickelate bilayer 
La$_3$Ni$_2$O$_7$ (LNO), we study a minimal mixed dimensional (MixD) $t-J$ model supplemented with a repulsive Coulomb interaction $V$. When hole-doped, previous numerical simulations revealed that the system exhibits strong binding energies, with a phenomenology resembling a BCS-to-BEC crossover accompanied by a Feshbach resonance between two distinct types of charge carriers. Here, we perform a mean-field analysis that enables a direct observation of the BCS-to-BEC crossover as well as microscopic insights into the crossover region and the pairing symmetry for two-dimensional bilayers. We benchmark our mean-field description by comparing it to density-matrix renormalization group (DMRG) simulations in quasi-one dimensional settings and find remarkably good agreement. For the two-dimensional system relevant to LNO our mean-field calculations predict a BCS pairing gap with an extended $s$-wave symmetry, directly resulting from the pairing mechanism's Feshbach-origin. Our analysis hence gives insights into pairing in unconventional superconductors and, further, can be tested in currently available ultracold atom experiments.
\end{abstract}

\maketitle



\begin{figure}
    \centering
    \includegraphics[width=0.8\linewidth]{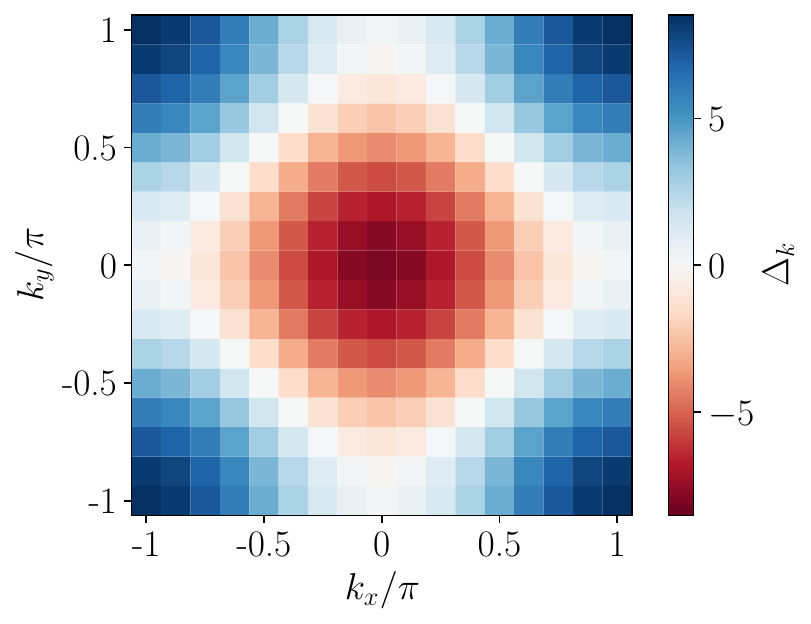}
    \caption{In the crossover regime, the spinon-chargon pairs are described by the BCS wavefunction. Here we consider a 2D MixD+V model on a $16 \times 16$ square lattice with $\frac{t_{\parallel}}{V}=0.1$, $J_{\perp}=V$ and $\delta = \frac{128}{256}$ holes in the system. The figure shows the value of the BCS order parameter $\Delta_k$ realizing an extended $s$-wave pairing symmetry.}
    \label{fig:extended_swave_pairing}
\end{figure}

Starting with the discovery of superconductivity in cuprate materials about four decades ago \cite{Bednorz1986,Lee2006,Scalapino1999,Nozieres:1985}, the search for superconductors with high critical temperatures has revealed unconventional superconductivity in various materials, including nickelate and cuprate compounds \cite{Schilling1993,Li2019}. Recently, the bilayer nickelate La$_3$Ni$_2$O$_7$ (LNO) was observed to become superconducting under pressure at a critical temperature of 80K \cite{Sun2023, zhang2023_zeroR, hou2023emergence}, with a pairing gap of extended $s$-wave symmetry predicted by theoretical works \cite{Zhang2024_electronic,Yang2023_possible,Liu2023_splusminus,Sakakibara2024_possible}, similar to those in infinite layer nickelates \cite{Kreisel2022_superconducting}, iron-based superconductors \cite{chubukov2011pairingmechanismfebasedsuperconductors}, and the strongly coupled bilayer Hubbard model \cite{Zhai20119_antiferro,Maier2011_pair,Nakata2017_finite}.
\begin{figure}[t]
    \centering
    \includegraphics[width=\linewidth]{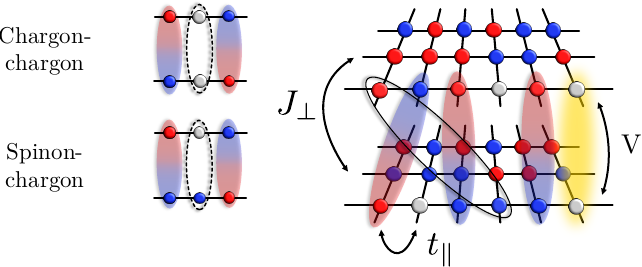}
    \caption{Schematic depiction of the 2D MixD+V bilayer model (right) and its quasi-particle constituents (left). We highlight the emergent constituents that characterize the two sides of the BCS-to-BEC crossover existing in this system: on the BCS side the system is dominated by spatially extended pairs ($sc^2$) of spinon-chargons ($sc$'s), involving one extended singlet (highlighted in the figure). The BEC side features tightly bound chargon-chargons ($cc$'s) characterized by the Coulomb repulsion $V$ (yellow shadow).}
    \label{fig:cartoon}
\end{figure}
Following the observation of high-T$_c$ superconductivity in LNO, theoretical studies suggested that a minimal effective low-energy model that governs the physics of these materials is a mixed dimensional (MixD) bilayer $t-J$ model, i.e., a single band $t-J$ model with suppressed inter-layer hopping but strong antiferromagnetic coupling between the layers (see Fig. \ref{fig:cartoon}) \cite{Luo2023, Lu2024Apr, oh2023type, qu2023}. More realistic multi-orbital models have also been proposed \cite{oh2023type, H.Yang23, yang2024strongpairingsymmetricpseudogap}. To find a minimal description for nickelate superconductors, investigating the pairing mechanism in MixD $t-J$ bilayers represents a crucial step toward developing a microscopic understanding of pairing in these unconventional superconductors. However, the nature of pairing and the pairing symmetry within this model remain elusive.  \\
\indent Since in contrast to conventional superconductors, high-$\text{T}_c$ superconductors feature significant pairing gaps, i.e. smaller Cooper pairs which possibly experience repulsive Coulomb interactions, the Coulomb repulsion $V$ is typically not negligible in these materials \cite{Lange2024Jan}. We account for that with an additional nearest neighbor repulsion $V$ between holes on the same rung of the bilayer. While this formulation is directly inspired by bilayer nickelates, the resulting MixD $t-J+V$ model also serves as a minimal platform to explore pairing in presence of strong correlations and short-range repulsion. Its mixed-dimensional structure and tunable interactions make it particularly suitable for implementation in ultracold atom systems, where key parameters can be engineered with high precision \cite{Hirthe2023Jan}. Thus, our choice of parameters is not meant as a literal fit for LNO but rather as an illustration of the pairing mechanism and pairing features of a minimal model in connection to the bilayer nickelate compounds.
These MixD+$V$ systems have been shown to host the following emergent structures upon doping the ground state at half-filling, consisting of singlets on each rung of the ladder \cite{Bohrdt2021Dec,Bohrdt2022Jun,Hirthe2023Jan,schlömer2023superconductivitypressurizednickelatela3ni2o7,Lange2024Jan,lange2024pairingdomeemergentfeshbach}: $(i)$ at small $V/J_\perp$ two holes form a tightly bound state of two chargons (chargon($c$)-chargon($c$) pair, $cc$'s) with strong binding energy. This is the result of inter-layer exchange $J_{\perp}$ and coherent movement of the $cc$'s through the system in order to minimize distortions of the (antiferromagnetic) AFM background \cite{Bohrdt2022Jun}. (ii) At large $V/J_\perp$ the system is dominated by spatially extended pairs of two partons, a spinon and a chargon (spinon($s$)-chargon($c$) pair, $sc$'s). In between $(i)$ and $(ii)$ a crossover from a Bose-Einstein condensate (BEC) of $cc$'s to a strongly correlated Bardeen–Cooper–Schrieffer (BCS) regime with pairs of $sc$'s ($sc^2$) is observed \cite{Lange2024Jan, H.Yang23}, which features a pairing dome with relatively high binding energy \cite{lange2024pairingdomeemergentfeshbach}. Each $sc^2$ is thus composed of two mesons ($sc$'s), comprising a total of four partons: two chargons and two spinons. \\
\indent While binding between holes in the $cc$'s can be explained with kinetic arguments, making it favorable for holes to move together through the system \cite{Bohrdt2021Dec,Bohrdt2022Jun}, binding of pairs of $sc$'s into $sc^2$ originates from a more subtle mechanism, leading to more spatially extended Cooper pairs of holes \cite{Lange2024Jan,lange2024pairingdomeemergentfeshbach}: in the BCS regime, where $sc$'s govern the low energy physics, numerical and analytical results suggest that binding is mediated via couplings to the high energy $cc$ channel, resembling a Feshbach resonance between the low energy (\enquote{open}) $sc$ channel and the high energy (\enquote{closed}) $cc$ channel. This Feshbach mediated pairing mechanism \cite{homeier2023, Crepel2021} features many similarities to the phenomenology of bilayer nickelate but also cuprate superconductors \cite{Lange2024Jan,lange2024pairingdomeemergentfeshbach,schlömer2023superconductivitypressurizednickelatela3ni2o7}: $(i)$ Pairing is facilitated by doping, leading to a dome of binding energies with its peak at intermediate doping; $(ii)$ Pairing only requires short-range antiferromagnetic (AFM) correlations but no long-range magnetic order; $(iii)$ The Fermi surface volume changes \cite{yang2024strongpairingsymmetricpseudogap}, similar to the small-to-large Fermi surface transition in cuprates; $(iv)$ For single layer models, the resulting pairing gap is predicted to have the $d$-wave symmetry observed in cuprates \cite{homeier2023}, and, as we will show here, for bilayers the extended $s$-wave symmetry predicted for LNO can be explained. So far, these observations have been made by comparing effective theories in terms of the emergent constituents and numerical observations in single and coupled ladders deep in the $cc$'s and $sc^2$ regimes. However, a description of the intermediate regime where $sc^2$ and $cc$ hybridize, needed to fully test the prediction of Feshbach mediated pairing, is lacking. \\
\indent Here, we go beyond earlier analysis and derive an effective mean-field description of the $sc$ and $cc$ constituents in the intermediate regime, that captures the essential physics of the BCS-to-BEC crossover and of the Feshbach resonance. We compare our mean-field predictions to density matrix renormalization group (DMRG) simulations of the MixD+$V$ ladder to validate the model. To further support the extension of the mean-field approach to higher dimensions we also perform the comparison for the quasi-2D two-ladder system, see Fig. \ref{fig:two-ladder}. Then, we apply our analysis to the 2D systems relevant to LNO. Our analysis allows to understand the BCS-to-BEC crossover in terms of the corresponding order parameters. Furthermore, it reveals the extended $s$-wave symmetry of the pairing gap as a direct consequence of the Feshbach mediated pairing mechanism, see Fig. \ref{fig:extended_swave_pairing}. \\
\section{Microscopic model}
The MixD $t-J$ bilayer supplemented with a repulsive interaction $V$ was first introduced in \cite{Lange2024Jan, lange2024pairingdomeemergentfeshbach, H.Yang23}. It features nearest neighbor hopping with amplitude $t_{\parallel}$, superexchange interaction $J_{\perp}$ and repulsive Coulomb interaction $V$ between holes on the same rung, i.e.
\begin{align} \label{Eq:microscopic model}
    \hat{\mathcal{H}} &= -              t_{\parallel}\hat{\mathcal{P}}\sum_{j}\sum_{\mu,\sigma}\left(\hat{c}_{j+1\mu\sigma}^{\dagger}\hat{c}_{j\mu\sigma}+\mathrm{h.c.}\right)\hat{\mathcal{P}} \notag \\ 
    &  + J_{\perp}\sum_{j}\left(\hat{\mathbf{S}}_{j0}\cdot\hat{\mathbf{S}}_{j1}-\frac{1}{4}\hat{n}_{j0}\hat{n}_{j1}\right)+V\sum_{j}\hat{n}_{j0}^{h}\hat{n}_{j1}^{h} \, 
\end{align} 
\begin{figure*}[ht]
  \centering
  \begin{tikzpicture}
    \node[inner sep=0] (imgA) {%
      \includegraphics[width=0.45\linewidth]{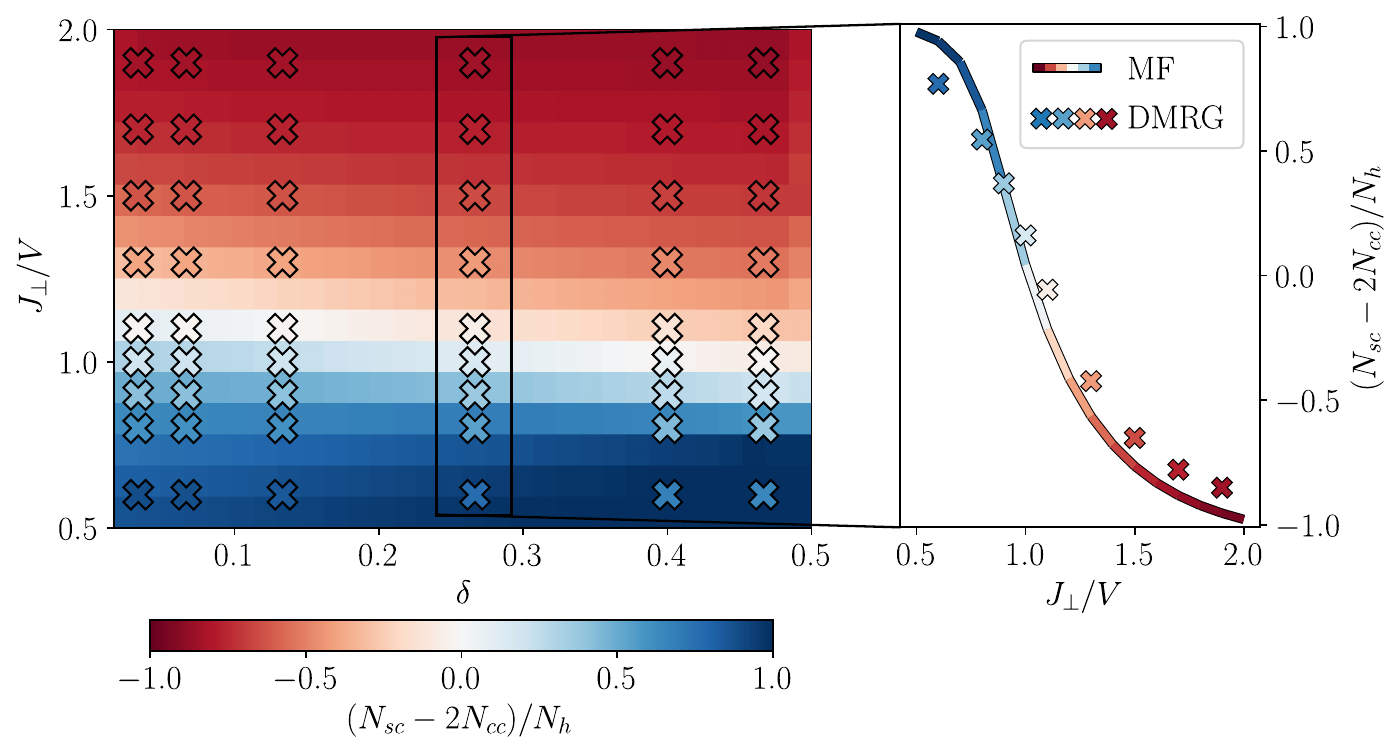}%
    };
    \node[
      anchor=north west,
      xshift=2pt, yshift=-7pt,
      font= \scriptsize
    ] at (imgA.north west) {(a)};
  \end{tikzpicture}%
  \quad
  \raisebox{6.5mm}{\begin{tikzpicture}
    \node[inner sep=0] (imgB) {%
      \includegraphics[width=0.45\linewidth]{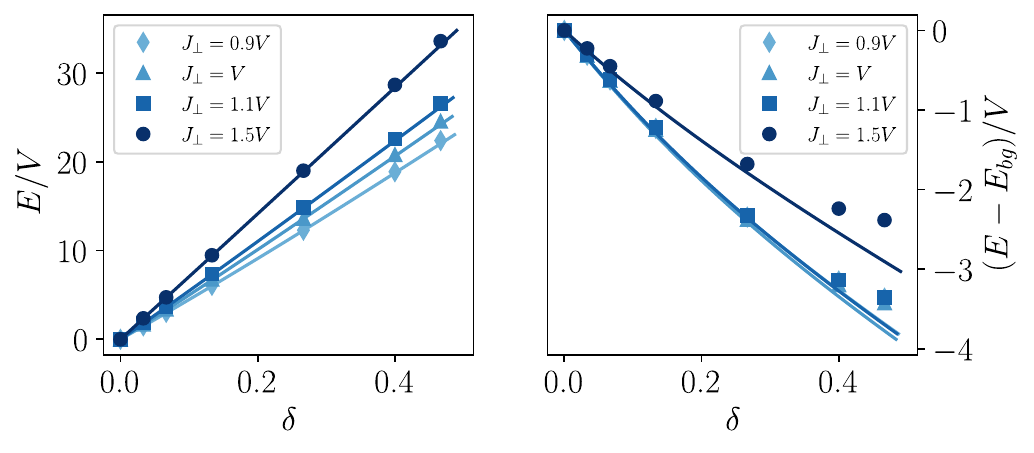}%
    };
    \node[
      anchor=north west,
      xshift=2pt, yshift=-2pt,
      font= \scriptsize
    ] at (imgB.north west) {(b)};
  \end{tikzpicture}
  }
  \caption{%
    Quasi-particle densities and ground state energy for the doped 1D ladder. (a) We compare the $sc$ and $cc$ densities for a 1D ladder as the ratio $\frac{J_{\perp}}{V}$ and the doping $\delta$ are tuned. We compute $(N_{sc}-2N_{cc})/N_h$ where $N_{sc}$ and $N_{cc}$ are respectively the number of $sc$'s and $cc$'s in the system as defined in Sec.~\ref{Sec:Benchmark}; $N_h$ is the total number of holes. 
    This allows to distinguish the two sides of the crossover. The inset shows the section at constant doping $\delta=\frac{8}{30}$ with the corresponding colors. We find good agreement between DMRG (crosses) and mean-field (colorplot/solid line) for a MixD+V ladder of size $L=30$ where we have fixed $\frac{t_\parallel}{V}=0.1$. 
    (b) The ground state energy $E$ (left) of the microscopic model for a 1D ladder system from DMRG (markers) is reproduced well by the mean-field prediction for the effective Hamiltonian (solid lines). The agreement is good also for the ground state energies without considering the contribution from the background distortion $E - E_{bg} = E-J_{\perp}N_{sc}-(V+ J_{\perp})N_{cc}$ (right).
  }
  \label{fig:1D-ladder-density/energy}
\end{figure*}
Here, $\hat{\mathcal{P}}$ is the Gutzwiller projector onto the subspace with maximum single occupancy per site. Spin and density operators at site $j$ and layer $\mu=0,1$ are represented by $\hat{\mathbf{S}}_{j\mu}$ and $\hat{n}_{j\mu} = \hat{n}_{j\mu\uparrow}+\hat{n}_{j\mu\downarrow}$. The hole density operators are denoted by $\hat{n}_{j\mu}^h = 1-\hat{n}_{j\mu}$.\\
\indent The motivation for studying the MixD+V model is twofold: firstly, the recently discovered bilayer nickelate $\text{La}_3\text{Ni}_2\text{O}_7$, showing superconducting behavior with $T_c = 80$K under high pressure, effectively realizes a bilayer MixD $t-J$ model. This can be seen from DFT calculations that reveal that the low energy degrees of freedom are the $3d_{z^2}$ and $3d_{x^2-y^2}$ Ni orbitals \cite{Lu2024Apr}, featuring strong Hund's coupling between them. The $d_{z^2}$ orbitals induce interlayer AFM superexchange interactions via the O-$2p$ orbitals in the intercalated LaO layer. In contrast, the $d_{x^2-y^2}$ orbitals are responsible for a weak intralayer hopping $t_{\parallel}$ and AFM coupling $J_{\parallel}$. In the limit of large Hund's coupling between the two $3d$ orbitals, the system realizes an effective AFM interaction $J_{\perp}$ between the two layers, and the interlayer hopping $t_{\perp}$ is suppressed \cite{Lu2024Apr}. Moreover, as stated before, we need to account for the Coulomb repulsion $V$ between neighboring holes, arising from the limited spatial extension of Cooper pairs in unconventional superconductors. The second motivation to study the MixD+$V$ model is their realizability in cold atom experiments: in Ref. \cite{Hirthe2023Jan}, a MixD ladder was realized by applying a potential offset between the legs of the ladder, effectively enhancing $J_\perp$ and suppressing the perpendicular hopping \cite{Hirthe2023Jan,Bohrdt2021Dec,Bohrdt2022Jun}. This system can be supplemented with nearest-neighbor repulsion by doping one leg with doublons and one with holes, see Refs. \cite{Lange2024Jan,lange2024pairingdomeemergentfeshbach,schlömer2024localcontrolmixeddimensions}. \\
\indent The ground state of this MixD+V model at zero doping consists of one singlet on each rung of the ladder \cite{Bohrdt2022Jun, Bohrdt2021Dec}. At finite doping, the physics of the system is determined by the competition between the kinetic energy of the holes and the energy from the distortion of the magnetic background due to the motion of the holes. In the tight-binding regime $t_{\parallel} \ll J_{\perp}, V$ the emerging constituents can be easily understood in the following two limits: when $J_{\perp}\gg V$, the low-energy constituents of the system are cc's, minimizing the distortion of the spin background. On the contrary, for $J_{\perp}\ll V$, each rung will tend to be occupied at maximum by one hole (and one spin), forming a spinon-chargon. In this regime, both emergent charge carriers can be assumed point-like, i.e., with constituents on the same rung \cite{Bohrdt2022Jun, Lange2024Jan}. In previous works \cite{Lange2024Jan, lange2024pairingdomeemergentfeshbach}, it was shown that the transition between the two limits can be explained as a crossover associated with a Feshbach resonance from tightly bound pairs of holes (closed channel) at small repulsion to more spatially extended, correlated pairs of individual holes (open channel) at large repulsion \cite{homeier2023}. In this picture, attractive interactions in the open channel are mediated by processes that couple to the closed channel even for large repulsion $V > J_{\perp}, t_{\parallel}$.
\\
\section{Mean-field theory}
Earlier studies \cite{Lange2024Jan, lange2024pairingdomeemergentfeshbach} provided a qualitative understanding of the physics of the MixD+$V$ model in the limiting cases where either $sc$'s (large $V/J_\perp$) or $cc$'s (small $V/J_\perp$) are dominant in the system, but the exploration of the crossover regime $J_{\perp} \approx V$ was limited to DMRG simulations of single and a few coupled ladders. Furthermore, such numerical methods do not directly provide access to the pairing gap. In contrast, here we obtain microscopic insight into the physics of the MixD+V model in the crossover regime where both $cc$'s and $sc$'s are present: we derive an effective two-channel Hamiltonian in terms of $sc$'s $f_{j\mu\sigma}^{(\dagger)}$ and $cc$'s $b_j^{(\dagger)}$ by performing a Schrieffer-Wolff transformation of Eq.(\ref{Eq:microscopic model}) assuming 
\begin{equation}
    t_{\parallel}\ll J_{\perp}, V 
\end{equation}
hence point-like $sc$'s and $cc$'s. Then, we develop a mean-field theory (MF) to solve the resulting model. \\
\indent In this limit, both $sc$'s and $cc$'s belong to the low-energy manifold, whereas the high-energy subspace contains configurations comprised of distorted singlets. The effective Hamiltonian we obtain reads
\begin{figure*}[t]
  \centering
  \begin{tikzpicture}
    \node[inner sep=0] (imgA) {%
      \includegraphics[width=0.45\linewidth]{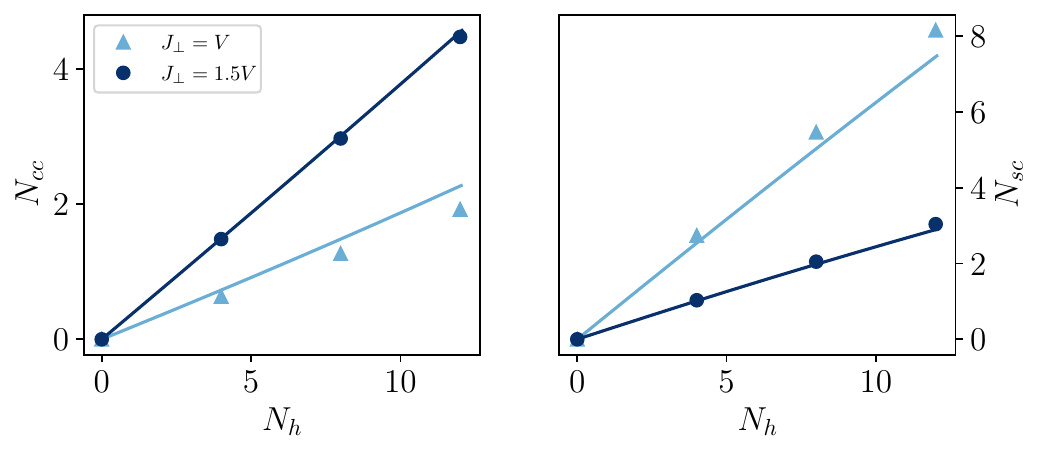}%
    };
    \node[
      anchor=north west,
      xshift=2pt, yshift=-2pt,
      font=\scriptsize
    ] at (imgA.north west) {(a)};
  \end{tikzpicture}%
  \quad
  \begin{tikzpicture}
    \node[inner sep=0] (imgB) {%
      \includegraphics[width=0.47\linewidth]{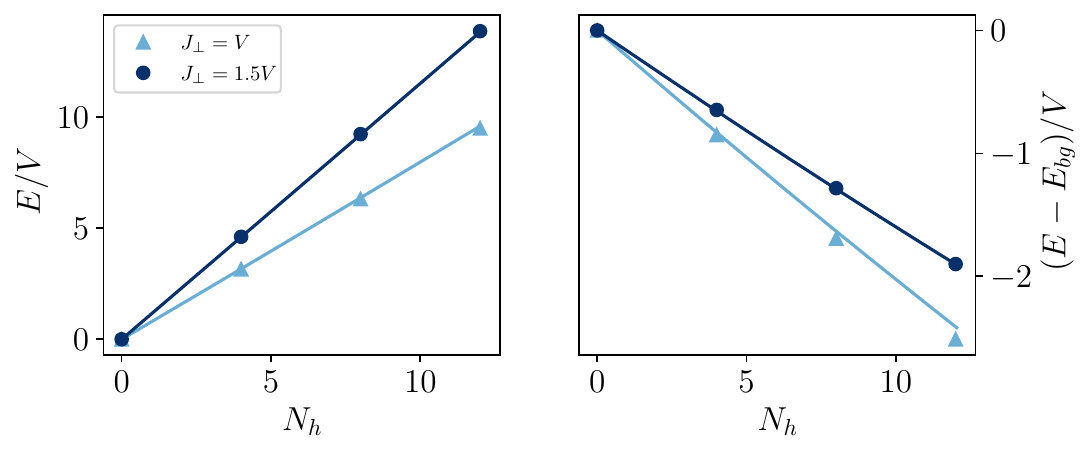}%
    };
    \node[
      anchor=north west,
      xshift=2pt, yshift=-2pt,
      font=\scriptsize
    ] at (imgB.north west) {(b)};
  \end{tikzpicture}
  \caption{
    Quasi-particle densities and ground state energy for the doped two-ladder system.(a) We compare the number of $sc$ and $cc$ pairs for a two‑ladder system with $L_x=30$ and $L_y=2$ as the number of holes is tuned. We show the DMRG (markers) and mean‑field (solid lines) for $J_\perp=V$ and $J_\perp=1.5V$, where $t_\parallel = 0.1V$ is fixed. 
    (b) The ground state energy $E$ (left) of the microscopic model for a two‑ladder system from DMRG (markers) is compared to the mean‑field prediction for the effective Hamiltonian (solid lines). We also show the results for the ground state energies without considering the contribution from the background distortion $E - E_{bg} = E - J_{\perp}N_{sc} - (V+J_{\perp})N_{cc}$ (right).
  }
  \label{fig:two-ladder}
\end{figure*}
\begin{align} \label{Eq:effective hamiltonian}
    \mathcal{\hat{H}}_{\text{eff}}^{sc,cc} &=
    \mathcal{\hat{H}}_{\rm kin}^{sc} + \mathcal{\hat{H}}_{\rm pot}^{sc} + \mathcal{\hat{H}}_{\rm pot}^{cc} + \mathcal{\hat{H}}_{\rm int}^{sc,sc} + \mathcal{\hat{H}}_{\rm rec}^{sc,cc} \\ \notag
    &+ \mu_h (\hat{N}_f + \hat{N}_b - N_h) 
\end{align}
\begin{align}\label{Eq:H_kin}
    \mathcal{\hat{H}}_{\rm kin}^{sc} &= + \tfrac{t_{\parallel}}{2} \sum_{\langle i,j \rangle} \sum_{\mu \sigma}  \mathcal{\hat{P}}_f \Big(\hat{f}^{\dagger}_{i \mu \sigma} \hat{f}_{j \mu \sigma} + h.c \Big)  \mathcal{\hat{P}}_f \\ \notag
    &- t_{\parallel} \sum_{\langle i,j \rangle} \sum_{\mu \sigma}  \mathcal{\hat{P}}_f \Big( \hat{f}_{i \mu \sigma}^{\dagger} \hat{b}_j^{\dagger} \hat{b}_i \hat{f}_{j \mu \sigma} + h.c. \Big)  \mathcal{\hat{P}}_f
\end{align}
\begin{equation}\label{Eq:H_pot_sc}
    \mathcal{\hat{H}}_{\rm pot}^{sc} = \Big(J_{\perp} - \tfrac{3}{2} \tfrac{t_{\parallel}^2}{J_{\perp}} \Big) \sum_{j \mu} \hat{n}_{j\mu}^f
\end{equation}
\begin{equation}\label{Eq:H_pot_cc}
    \mathcal{\hat{H}}_{\rm pot}^{cc} = \big( V+J_{\perp} \big) \sum_{i} \hat{b}_i^{\dagger} \hat{b}_i
\end{equation}
\begin{align}\label{Eq:H_int_sc}
    \mathcal{\hat{H}}_{\rm int}^{sc,sc} &= + \tfrac{3}{2} \tfrac{t_{\parallel}^2}{J_{\perp}} \sum_{\langle i,j \rangle} \sum_{\mu \mu'} \hat{n}_{i\mu}^f \hat{n}_{j\mu'}^f \\ \notag  
    & - 4 \tfrac{t^2_{\parallel}}{V} \sum_{\langle i,j \rangle} \Big( \boldsymbol{-} \mathbf{\hat{J}}_i \cdot \mathbf{\hat{J}}_j + \tfrac{1}{4} \Big) \Big( \mathbf{\hat{S}}_i \cdot \mathbf{\hat{S}}_j + \tfrac{3}{4} \hat{n}_i^f \hat{n}_j^f \Big)
\end{align}
\begin{equation}\label{Eq:H_rec}
   \mathcal{\hat{H}}_{\rm rec}^{sc,cc} = - \frac{t_{\parallel}}{\sqrt{2}} \sum_{\langle i,j \rangle} \sum_{\mu, \sigma} (-1)^{\sigma}  \mathcal{\hat{P}}_f \big(\hat{f}^{\dagger}_{i \mu \sigma}  \hat{f}^{\dagger}_{j \thickbar{\mu} \thickbar{\sigma}} \hat{b}_j + h. c. \big)  \mathcal{\hat{P}}_f 
\end{equation}
where Eq.(\ref{Eq:H_kin}) corresponds to the kinetic energy of the $sc$'s and $cc$'s, and Eq.(\ref{Eq:H_pot_sc}, \ref{Eq:H_pot_cc}) are respectively the terms associated with the $sc$'s and the $cc$'s chemical potential.  The term in Eq.(\ref{Eq:H_int_sc}) takes into account the interactions between $sc$'s sitting on neighboring rungs, which are subject to repulsive interaction $\propto + \tfrac{3}{2} \tfrac{t_{\parallel}^2}{J_{\perp}}$ and attractive interaction $\propto \tfrac{t_{\parallel}^2}{V}$ for triplet states. The term in Eq.(\ref{Eq:H_rec}) describes the recombination of a pair of two neighboring $sc$'s (open channel) into one $cc$'s (closed channel) and vice versa
giving rise to the Feshbach resonance as the presence of $cc$'s lowers the energy of the $sc$ channel. The last term $\propto \mu_h$ fixes the total number of holes in the system. In the above, $\hat{n}_{j\mu}^f = \hat{n}_{j\mu\uparrow}^f + \hat{n}_{j\mu\downarrow}^f$ denotes the $sc$'s density while the layer-isospin operator $\mathbf{\hat{J}_i}$ and the spin operator $\mathbf{\hat{S}_i}$ are defined in Appendix \ref{Appendix: Eff H}. 
\indent In order to capture the low-energy physics of $(sc)^2$-to-$cc$ crossover we consider the following ansatz for the ground state wavefunction
\begin{align} \label{Eq:ansatz wavefunction}
    \ket{\Psi} &= \bigg[ \prod_k \Big(u_k + v_k \frac{1}{\sqrt{2}} \sum_{\sigma} (-1)^{\sigma} \hat{f}^{\dagger}_{k\mu\sigma} \hat{f}^{\dagger}_{-k\thickbar{\mu}\thickbar{\sigma}} \Big) \notag \\
    & \otimes \mathcal{N} \exp\big({\beta \hat{b}_{k=0}^{\dagger}} \big) \bigg] \ket{0}
\end{align}
\noindent 
where the first term accounts for the BCS pairing of $sc$'s with variational parameters $u_k, v_k \in \mathbb{C}$. The explicit singlet structure of the $sc$'s (note that $\thickbar{\sigma}$ and $\thickbar{\mu}$ denote the respective opposite spin and layer) is motivated by the fact that for sufficiently large $V$  holes on the same rung strongly repel each other due to Coulomb interaction. Therefore as $\frac{J_\perp}{V}$ is decreased, the $cc$'s are disrupted and the displacement of the two holes consituting the $cc$ leads to the formation of a tilted singlet between two neighbouring spinons on opposite layers. The second term describes the BEC of tightly bound $cc$'s with normalization constant $\mathcal{N}$ and order parameter $\beta \in \mathbb{C}$. Here, $\ket{0}$ denotes the $sc$'s and $cc$'s vacuum state, consisting of singlets on each rung. Our goal in the following is to tune $\frac{J_\perp}{V}$ to transition from the BEC regime of tightly bound $cc$'s, where we expect large $\beta$, to the BCS regime where the $sc$'s form extended Cooper pairs in a BCS state. At the resonance, between the two regimes, the effective $sc$ scattering length diverges. To describe this intermediate regime, it is important to include explicitly the coupling between the two channels ($sc$ and $cc$). \\
\indent By minimizing the energy expectation value of the Hamiltonian in Eq.(\ref{Eq:effective hamiltonian}) with respect to the variational parameters of the wavefunction in Eq.(\ref{Eq:ansatz wavefunction}), $\partial_{u_k v_k} \partial_{\beta} \langle \Psi | \mathcal{\hat{H}}_{\text{eff}}^{sc,cc} | \Psi \rangle = 0$, see Appendix \ref{appendix: gs parameters} (including Refs. \cite{Parish_2014}), we obtain the ground state properties of the system. In order to analytically perform the mean-field calculations we neglect the Gutzwiller projections in Eq.(\ref{Eq:effective hamiltonian}). This is expected to be a good approximation for low doping and we restrict our analysis to values of the hole doping $\delta<0.5$. \\

\section{Benchmarking Mean-Field in 1D MixD+V ladders} \label{Sec:Benchmark}
\begin{figure}[t!] 
    \centering
    \includegraphics[width=0.9\linewidth]{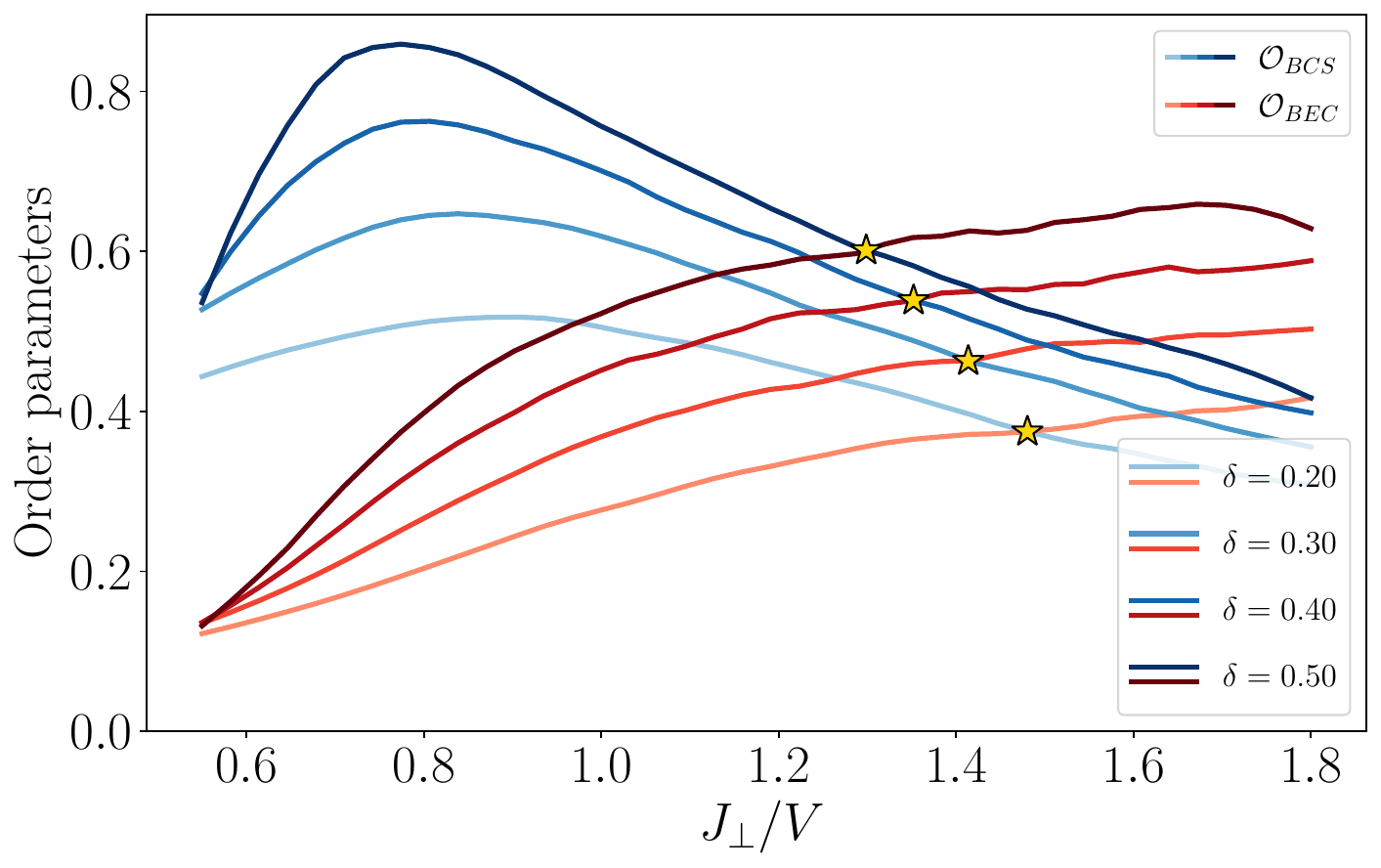}
    \caption{We show the order parameters for the BEC and BCS regimes of the ansatz wavefunction in a 2D bilayer system, as defined in the main text. Here we consider a system of size $L_x = L_y = 16$ for doping $\delta = \{\frac{38}{256},\frac{64}{256}, \frac{89}{256}, \frac{115}{256} \}$ at fixed $t_\parallel=0.1V$. The stars mark the crossing point for each doping level.}
    \label{fig:order parameters}
\end{figure}

We compare the results from our mean-field analysis with DMRG simulations, obtained using the package SyTen with particle conservation in each leg and a global $U(1)_{S_z}$ symmetry \cite{Hubig2017Oct, syten, syten1, syten2, Schloemer2023}. In Fig. \ref{fig:1D-ladder-density/energy} we focus on 1D ladder systems where the $sc$'s and $cc$'s densities of the microscopic model of Eq.(\ref{Eq:microscopic model}) computed with DMRG are compared to the ground state mean-field results obtained from the effective Hamiltonian in Eq.(\ref{Eq:effective hamiltonian}). In particular, the difference between the number of holes contributing to the $sc$'s sector and $cc$'s sector $\propto N_{sc}-2N_{cc}$ is shown as a function of $\tfrac{J_{\perp}}{V}$ and hole doping $\delta$. The two different regimes can be identified by the change in sign of the above quantity. Here, $\hat{n}_{cc} = \frac{1}{L} \sum_i \hat{n}_{i\mu=1}^h \hat{n}_{i\mu=0}^h$ and $\hat{n}_{sc} = \frac{1}{L} \sum_i \hat{n}_{i\mu=1}^h (1- \hat{n}_{i\mu=0}^h)$ for the microscopic model, and $\hat{n}_{cc} = \frac{1}{L} \sum_i \hat{b}_i^{\dagger} \hat{b}_i$ and $\hat{n}_{sc} = \frac{1}{L} \sum_{i\mu} \hat{n}^f_{i\mu}$  for the mean-field (MF) theory. As for the MF calculations, both species are considered to be point-like since the system is probed in the tight binding regime $t_{\parallel}\ll J_{\perp},V$. We find that for $V > J_{\perp}$ the system is dominated by $sc$'s and the number of $cc$'s is negligible. For lower doping however, it is expected to grow linearly for $\delta > 0.5$ due to the single occupancy restriction \cite{Lange2024Jan}, which we cannot capture within MF theory. For $V < J_{\perp}$ the formation of $sc$'s is suppressed and the holes are bound together in $cc$'s. 
\indent Fig. \ref{fig:1D-ladder-density/energy}a compares the ground state energy obtained from DMRG and MF theory for different values of $\frac{J_{\perp}}{V}$ at low doping. In the tight binding regime,  when the system is hole doped, a large contribution to the ground state energy comes from the distortion of the AFM background: $J_{\perp}N_{sc}$ for the $sc$'s and $(V+ J_{\perp})N_{cc}$ for the $cc$'s. However, the role of the kinetic, interaction, and recombination terms is not negligible; this is shown in Fig. \ref{fig:1D-ladder-density/energy}b where we compare the ground state energies after subtracting the $sc$'s and $cc$'s background terms mentioned above. We also extend the analysis to the quasi-2D two-ladder system and report the results in Fig. \ref{fig:two-ladder}. In both cases, there is good agreement, validating our MF approach. \\
\indent Next, we calculate the binding energies, defined as 
\begin{equation}\label{eq:binding energy}
    E_B(N_h) = 2(E_{N_h-2} - E_{N_h-1}) - (E_{N_h-2}-E_{N_h})
\end{equation}
\begin{figure}[t!]
    \centering
    \includegraphics[width=0.8\linewidth]{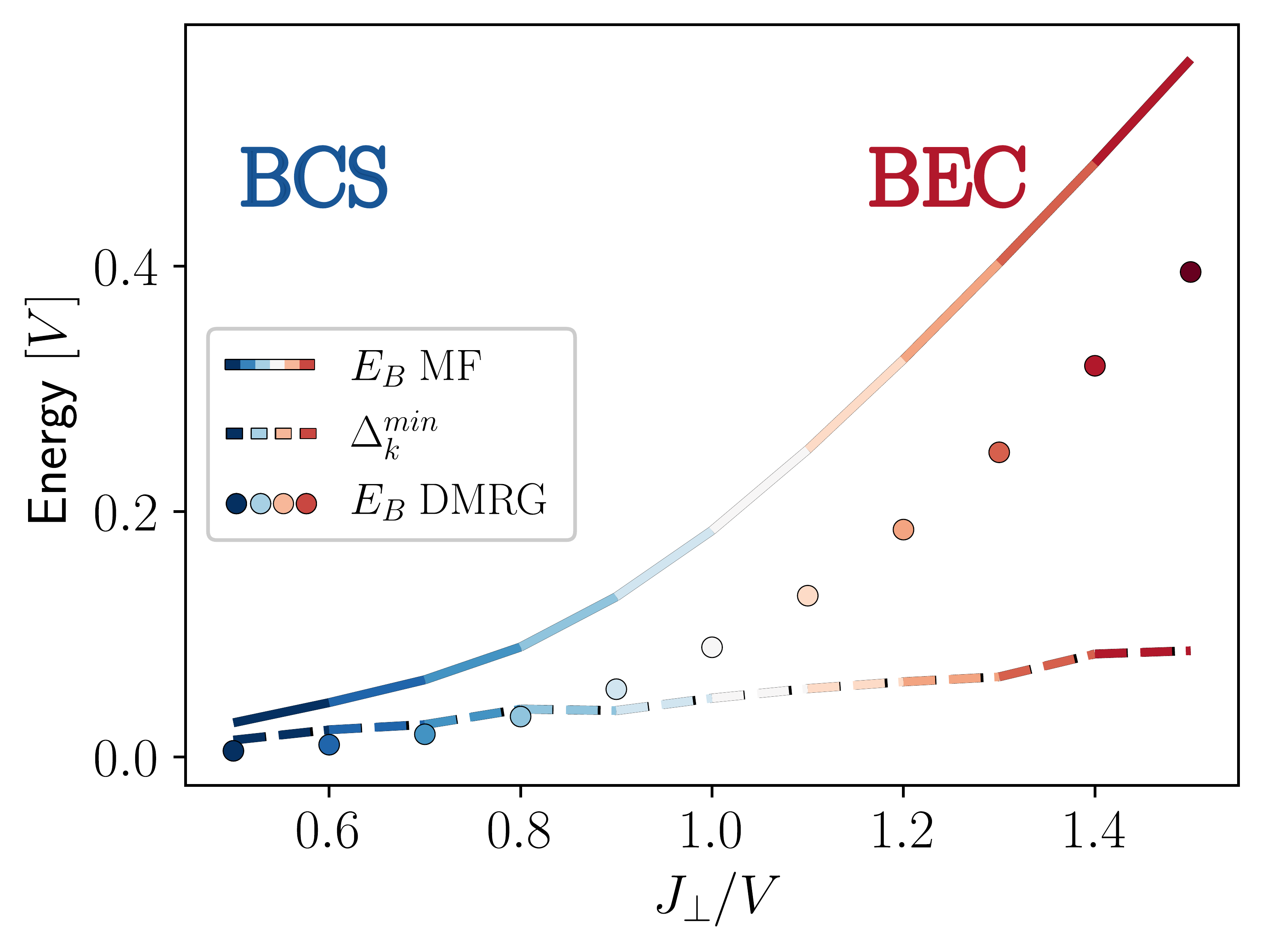}
    \caption{The binding energies $E_B$ computed for a 1D ladder system via DMRG (markers) is compared to mean-field theory (solid line) as $J_{\perp}$ is tuned. Here, we also show the value of the order parameter $\Delta_k$ at the minimum of $E_k$, i.e. where the band gap is minimal (dashed line). We consider a ladder system of size $L=30$ with fixed doping $\delta=\frac{14}{30}$ and $\frac{t_\parallel}{V}=0.1$.}
    \label{fig:binding}
\end{figure} 
\noindent where $N_h$ is the number of holes in the system, added alternatingly to each layer. The expression can be evaluated directly from the DMRG ground state energies at different doping. For the mean-field theory, we obtain the binding energy by computing the terms in Eq.\ref{eq:binding energy}) separately. We consider the system in the BEC regime; here, the addition of two holes leads to the formation of a $cc$ and $E_{N_{h}-2}-E_{N_h} \approx 0$ since the energy required to add a quasi-particle to the condensate is approximately zero. However, the addition of one single hole corresponds to the creation of a single quasi-particle in the BCS state which leads to an energy variation of $E_{N_{h}-1}-E_{N_h} = \min_k(E_k)$ with $E_k = \sqrt{\Delta_k^2 + \xi_k^2}$ where $\Delta_k$ is the BCS order parameter and $\xi_k$ is the dispersion relation for $sc$'s, see Appendix \ref{appendix: gs parameters}. Therefore the binding energies are given by  $E_B = 2(E_{N_h-2} - E_{N_h-1}) - (E_{N_h-2}-E_{N_h}) \approx 2(\min_k(E_k)) - 0 = 2\min_k(E_k)$ , which corresponds to the pairing gap obtained by diagonalizing the BCS mean-field Hamiltonian, see Appendix \ref{Appendix: binding E}. As shown in Fig. \ref{fig:binding} we find that as $\frac{J_{\perp}}{V}$ increases, the binding energy $E_B$ becomes larger, in agreement with the picture of tightly bound pairs of holes \cite{Bohrdt2022Jun, Hirthe2023Jan}. The values obtained from the mean-field calculations qualitatively follow the results from DMRG simulations.
\indent Another indication that the system undergoes a BCS-to-BEC crossover is obtained by comparing the binding energy $E_B$ of the fermion pairs with the order parameter $\Delta_k$ evaluated at $k^*$ where the band gap is minimal $\Delta_k^{\rm min} = \Delta_k(k^*)$ where $k^* = $ arg min($E_k$). This accounts for the phase coherence between the pairs: in the BCS region of extended pairs the two quantities converge since pairs form and develop phase coherence at the same energy, whereas they are significantly different in the BEC regime, where fermions pair up with binding energy $E_B$ but the condensation energy is lower \cite{SadeMelo1993Nov, SadeMelo2008Oct}. This is a hallmark of the BCS-to-BEC crossover. \\
\indent Furthermore, the mean-field description provides direct access to the order parameters characterizing the BEC and BCS regimes. A key requirement for these order parameters is that they remain well-defined in the thermodynamic limit, $L \rightarrow \infty$. To this end, we define
\begin{align} 
    \mathcal{O}_{BEC} &= \sqrt{\frac{N_{cc}}{L^d}},\\ 
    \mathcal{O}_{BCS} &= \frac{1}{L^d} \sum_k \frac{|\Delta_k|}{\sqrt{\Delta_k^2+\xi_k^2}} \\
    &\xrightarrow[]{L \rightarrow \infty} \frac{1}{(2 \pi)^d} \int_{\text{BZ}} d^d k \frac{|\Delta_k|}{\sqrt{\Delta_k^2+\xi_k^2}}, \notag
\end{align} 
which correspond to the BEC and BCS order parameters, respectively, for a $d$-dimensional system. The summation over the Brillouin zone (BZ) ensures that the order parameter is defined without resolving individual momentum states in $k$-space. In Fig. \ref{fig:order parameters} we analyze a 2D bilayer system $d=2$. This analysis confirms our picture of a BCS-to-BEC crossover around $\frac{J_\perp}{V}\approx 1.4$.
\section{Extended s-wave pairing in 2D MixD+V models}
In the previous section, we compared the mean-field predictions for different observables with the DMRG simulations of the microscopic model in 1D. This showed that the MixD+V ladder under consideration realizes a BCS-to-BEC crossover as $\frac{J_{\perp}}{V}$ and $\delta$ are tuned. The similarly good agreement in the quasi‑2D ladder system in Fig. \ref{fig:two-ladder} indicates that our mean-field approach can be extended to true 2D bilayer systems, thereby granting access to the pairing gap. Indeed, the $sc$'s in the system pair up in singlets on opposite layers and are effectively described by the BCS wavefunction in Eq.(\ref{Eq:ansatz wavefunction}), which is characterized by the parameter $\Delta_k$, defined in Appendix \ref{appendix: gs parameters}. Figure \ref{fig:order parameters} shows that the 2D crossover closely parallels the 1D case: as the interlayer AFM coupling \(J_{\perp}\) grows, extended \(sc\) pairs are replaced by tightly bound \(cc\) pairs, marking a BCS-BEC transition. Increasing the doping from 0.2 to 0.5 shifts this crossover to lower \(\tfrac{J_{\perp}}{V} \in (1.2, 1.5)\), consistent with the quasi‑particle densities in Fig. \ref{fig:1D-ladder-density/energy}a, which likewise show earlier \(cc\) dominance at higher doping. Although this is a smooth crossover rather than a sharp phase transition, its midpoint provides a useful indicator of the underlying shift in pairing character. Moreover, as shown in Fig. \ref{fig:extended_swave_pairing}, the pairing function associated with the $(sc)^2$ in a bilayer square lattice system features an extended $s$-wave symmetry, i.e. it presents a symmetric non trivial momentum dependence with a clear nodal region and sign inversion.\\
\begin{figure}[t!]
    \centering
    \includegraphics[width=0.8\linewidth]{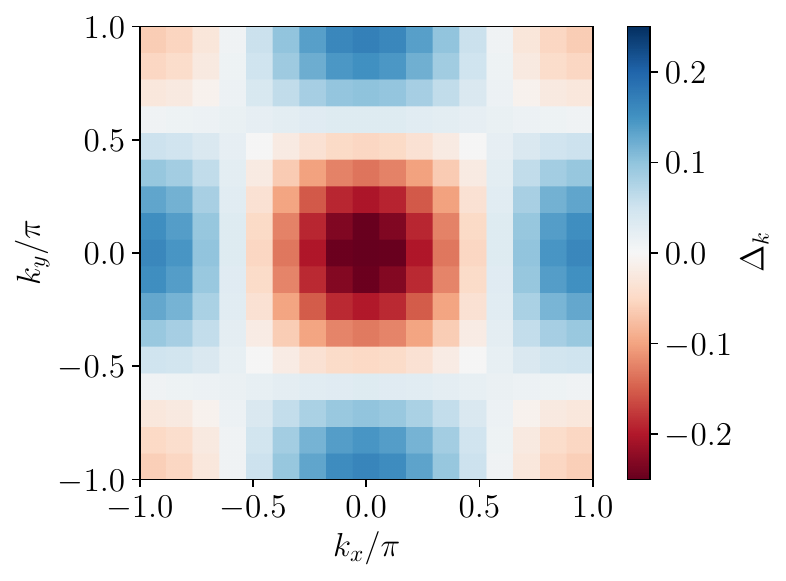}
    \caption{The pairing gap $\Delta_k$ is modified by the introduction of diagonal recombinations of sc's and cc's and an additional region with opposite sign appears at the corners of the Brillouin zone. Here we consider a 2D MixD+V model on a 16 × 16 square lattice with $\frac{t_\parallel}{V}=0.1$, $J_{\perp} = V$, $\delta = \frac{104}{256}$ holes in the system, and with $\epsilon = 1.5$.}
    \label{fig:diagonal_pairing}
\end{figure}
\indent This prediction is particularly intriguing in the light of the results form Ref. \cite{Yang2023_possible} where Cooper pairing with $s_{\pm}$-wave symmetry was found using functional renormalization group techniques. We note that recent two‑orbital calculations for La\(_3\)Ni\(_2\)O\(_7\) show that the superconducting gap symmetry is highly sensitive to the Ni-\(e_g\) crystal‑field splitting \(\Delta_{\rm CF}\): tuning \(\Delta_{\rm CF}\) alters the relative strength of intra‑ and interlayer superexchange and can drive a transition from \(d_{xy}\) to \(s_\pm\) pairing \cite{xia_liu_zhou_chen_2025}. The pairing gap that we obtain in our mean-field analysis has a simple structure and shows only two distinct areas in the Brillouin zone with different sign. We argue that this is due to the assumption of point-like $sc$'s and $cc$'s which is valid in the tight binding regime $t_{\parallel} \ll V,J_{\perp}$. However, additional features of the pairing gap are expected to appear when considering the internal structure of the constituents which becomes relevant for larger $t_{\parallel}$. Specifically, allowing recombination of $sc$'s and $cc$'s along the diagonal directions of the lattice, leads to the formation of a third region at the corners of the Brillouin zone, see Fig. \ref{fig:diagonal_pairing}. The diagonal recombination is implemented in the Hamiltonian of Eq. (\ref{Eq:effective hamiltonian}) by introducing the term 
\begin{equation}
    \mathcal{\hat{H}}_{\rm diag \, \rm rec}^{sc,cc} = - \epsilon \frac{t_{\parallel}}{\sqrt{2}} \sum_{i,j} \sum_{\mu, \sigma} (-1)^{\sigma}  \mathcal{\hat{P}}_f \big(\hat{f}^{\dagger}_{i \mu \sigma}  \hat{f}^{\dagger}_{j \thickbar{\mu} \thickbar{\sigma}} \hat{b}_j + h. c. \big)  \mathcal{\hat{P}}_f
\end{equation}
where $\epsilon$ tunes the strength of the interaction between the two channels along the $i,j$ diagonal neighbors.  \\
\section{Conclusion}
In conclusion, we have explored the physics of the MixD+V bilayer system and its implications for high-$\text{T}_{\text{c}}$ superconductivity in the bilayer nickelate La$_3$Ni$_2$O$_7$ (LNO). Previous work \cite{Bohrdt2022Jun, Lange2024Jan, lange2024pairingdomeemergentfeshbach, H.Yang23, Lu2024Apr, qu2023, oh2023type} identified two different regimes upon hole doping the effective low energy model of these materials: on one side, the BCS region of spatially extended $sc$'s Cooper pairs, and on the other, the BEC region of tightly bound $cc$'s. Additionally, our numerical analysis indicates that in the BCS limit, pairing between $sc$'s arises from a Feshbach resonance, in analogy with atomic physics. However, a microscopic understanding of the intermediate regime was lacking. In this study, starting from the microscopic model, we derived an effective Hamiltonian in terms of the $sc$'s and $cc$'s, and proposed an ansatz ground state wavefunction to capture the essential physics of the charge carriers across the BCS-to-BEC crossover. The formulation of the mean-field theory allowed us to gain a microscopic understanding of the system by constructing a clear conceptual framework where the emergent constituents and their interactions could be investigated. By comparison with DMRG simulations in ladder systems we have demonstrated the validity of the model and showed that it effectively describes the low energy physics of the system. Notably, the main result of this work is the prediction of a pairing gap with extended $s$-wave symmetry in the bilayer model, which characterizes the $(sc)^2$ Cooper pairs. Our finding paves the way to explore even more complex scenarios by considering the internal structure of the charge carriers, which we propose will lead to additional features in the pairing gap, see Fig. \ref{fig:diagonal_pairing}. Lastly, we remark that the MixD+V bilayer system can be studied using ultracold atom experiments by a modification of the setup in Ref. \cite{Hirthe2023Jan}, incorporating nearest-neighbor repulsion \cite{Lange2024Jan}.
\section{Acknowledgments}
We would like to thank Annabelle Bohrdt, Lukas Homeier, Henning Schlömer and Matjaz Kebric for helpful discussions. We acknowledge funding by the Deutsche Forschungsgemeinschaft (DFG, German Research Foundation) under Germany's Excellence Strategy -- EXC-2111 -- 390814868 and from the European Research Council (ERC) under the European Union’s Horizon 2020 research and innovation programm (Grant Agreement no 948141) — ERC Starting Grant SimUcQuam. HL acknowledges support by the International Max Planck Research School.


\bibliography{bibsmap}


\onecolumngrid
\newpage
\appendix
\section{Derivation of the effective Hamiltonian} \label{Appendix: Eff H}
\noindent
In this section, we derive the effective Hamiltonian describing the system in terms of spinon-chargons (sc) and chargon-chargons (cc); subsequently, we show how to obtain the ground-state wavefunction parameters through the variational method.
Our study focuses on the intermediate regime in the crossover region where $t_{\parallel} \ll J_{\perp} \simeq V $. In order to derive the effective spinon-chargon and chargon-chargon Hamiltonian from the microscopic MixD Hamiltonian, we perform Schrieffer-Wolff transformations. \\
Here, we use the notation introduced previously for the spinon-chargon and chargon-chargon annihilation (creation) operators: $f_{j\mu\sigma}^{(\dagger)}$ and $b_j^{(\dagger)}$ \cite{Lange2024Jan}
\begin{equation}
    \hat{f}^{\dagger}_{i\mu\sigma} \ket{\ldots \singlet \ldots} = \ket{ \ldots \scup \ldots}
\end{equation}
\begin{equation}
    \hat{f}_{i\mu\sigma} \ket{\ldots \scup \ldots} = \ket{\ldots \singlet \ldots}
\end{equation}
\noindent with $\mu \in \{0,1\} $ denoting the leg degree of freedom of the spinon-chargon, and
\begin{equation}
    \hat{b}_i^{\dagger} \ket{\ldots \singlet \ldots } = \ket{\ldots \cc \ldots} 
\end{equation}
\begin{equation}
     \hat{b}_i \ket{\ldots \cc \ldots } =  \ket{\ldots \singlet \ldots }
\end{equation} 
\noindent where singlets are identified by \singlet, chargon-chargon pairs by \cc and where \scup and \scdown denote spinon-chargon pairs in the upper and lower layer, respectively. The sc-cc vacuum consists of singlets on each rung, represented by $\ket{\ldots \singlet \ldots}$. 
Isolated spinon-chargons allow low-energy (Gutwiller projected) hopping processes \cite{Lange2024Jan}
\begin{equation}
    \mathcal{\hat{H}}_{sc}^{(1)} = - \frac{t_{\parallel}}{2} \sum_{\langle i,j \rangle} \sum_{\mu \sigma} \mathcal{\hat{P}}_f \Big(\hat{f}^{\dagger}_{i \mu \sigma} \hat{f}_{j \mu \sigma} + \text{h.c} \Big) \mathcal{\hat{P}}_f
\end{equation}
\noindent In this regime, recombination processes involving chargon-chargon and spinon-chargon pairs remain in the low-energy channel
\begin{equation}\label{cc sc singlet}
    \mathcal{\hat{H}}_{cc,sc}^{(1)} = - \frac{t_{\parallel}}{\sqrt{2}} \sum_{\langle i,j \rangle} \sum_{\mu \sigma} (-1)^{\sigma} \big(\hat{f}^{\dagger}_{i \mu \sigma}  \hat{f}^{\dagger}_{j \thickbar{\mu} \thickbar{\sigma}} \hat{b}_j + h. c. \big)
\end{equation}
\noindent
For isolated spinon-chargons, we also have to consider second-order processes due to recombination with the high-energy channel of tilted singlets, which have an amplitude of $-2\tfrac{3}{4}\tfrac{t_{\parallel}^2}{J_{\perp}}$ where the factor of 2 comes from the two directions in which the particles can hop \cite{Lange2024Jan}. Moreover, each spinon-chargon contributes $+J_{\perp}$ for every broken singlet w.r.t. the sc-cc vacuum. Together it gives
\begin{equation} \label{Appendix:H_sc^2}
   \hat{\mathcal{H}}_{sc}^{(2)} = \Big(J_{\perp} - \frac{3}{2} \frac{t_{\parallel}^2}{J_{\perp}} \Big) \sum_{j \mu} \hat{n}_{j\mu}^f 
\end{equation}
\noindent
If two spinon-chargons occupy neighboring rungs, second-order recombination in each other's direction is obstructed, and we need to take that into account by introducing the additional term
\begin{equation}
    +\frac{3}{2} \frac{t_{\parallel}^2}{J_{\perp}} \sum_{\langle i,j \rangle} \sum_{\mu \mu'} \hat{n}_{i\mu}^f \hat{n}_{j\mu'}^f 
\end{equation}
\noindent When spinon-chargons and chargon-chargons are located on neighboring rungs, they can exchange position by the first-order hopping process of one particle
\begin{equation}
    \mathcal{\hat{H}}_{sc,cc}^{(1)} = -t_{\parallel} \sum_{\langle i,j \rangle} \sum_{\mu \sigma} \Big( \hat{f}_{i \mu \sigma}^{\dagger} \hat{b}_j^{\dagger} \hat{b}_i \hat{f}_{j \mu \sigma} + h.c. \Big)
\end{equation}
\noindent Neighboring spinon-chargons can recombine via the high-energy triplet channel V in a second-order process
\begin{align}
    \mathcal{\hat{H}}_{\rm triplet} =& 
     \, 2 \frac{t^2_{\parallel}}{V} \sum_{\langle i,j \rangle} \Big(J^{+}_i J^{-}_j + J^{-}_i J^{+}_j + 2 J^z_i J^z_j -\frac{1}{2} \Big)  \Big( \mathbf{\hat{S}}_i \cdot \mathbf{\hat{S}}_j + \frac{3}{4} \hat{n}_i^f \hat{n}_j^f \Big) \\
    = & -4 \frac{t^2_{\parallel}}{V} \sum_{\langle i,j \rangle} \Big( \boldsymbol{-} \mathbf{\hat{J}}_i \cdot \mathbf{\hat{J}}_j + \frac{1}{4} \Big) \Big( \mathbf{\hat{S}}_i \cdot \mathbf{\hat{S}}_j + \frac{3}{4} \hat{n}_i^f \hat{n}_j^f \Big)
\end{align}
\noindent
where $\hat{\textbf{J}}_i = \frac{1}{2} \sum_{\sigma} \sum_{\mu \mu'} \hat{f}_{i \mu \sigma}^{\dagger} \boldsymbol{\sigma}_{\mu \mu'} \hat{f}_{i \mu' \sigma}$ is the layer isospin operator and $\hat{P}_T = \Big( \mathbf{\hat{S}}_i \cdot \mathbf{\hat{S}}_j + \frac{3}{4} \hat{n}_i^f \hat{n}_j^f \Big)$ is the projector onto the triplet state with $\hat{\mathbf{S}}_i = \frac{1}{2} \sum_{\mu} \sum_{\sigma \sigma'} \hat{f}_{i \mu \sigma}^{\dagger} \boldsymbol{\sigma}_{\sigma \sigma'} \hat{f}_{i \mu \sigma}$.
\noindent The singlet channel lies in the low-energy manifold and has already been included in Eq. (\ref{cc sc singlet}).
Lastly, we need to include the energy for the chargon-chargons in the system with respect to the vacuum state: for each cc, we have a broken singlet, which contributes $+J_{\perp}$ and the on-rung interaction $V$
\begin{equation}
    \big(V+J_{\perp} \big) \sum_{i} \hat{b}_i^{\dagger} \hat{b}_i 
\end{equation}
\noindent
The full-perturbative Hamiltonian is
\begin{align} \label{Appendix:HAmiltonian real space}
    \mathcal{\hat{H}}_{\text{eff}}^{sc,cc} &=  
     - \frac{t_{\parallel}}{\sqrt{2}} \sum_{\langle i,j \rangle} \sum_{\mu \sigma} (-1)^{\sigma}  \mathcal{\hat{P}}_f \big(\hat{f}^{\dagger}_{i \mu \sigma}  \hat{f}^{\dagger}_{j \thickbar{\mu} \thickbar{\sigma}} \hat{b}_j + h. c. \big)  \mathcal{\hat{P}}_f
     + \frac{t_{\parallel}}{2} \sum_{\langle i,j \rangle} \sum_{\mu \sigma}  \mathcal{\hat{P}}_f \Big(\hat{f}^{\dagger}_{i \mu \sigma} \hat{f}_{j \mu \sigma} + h.c \Big)  \mathcal{\hat{P}}_f \notag\\
    &\quad- t_{\parallel} \sum_{\langle i,j \rangle} \sum_{\mu \sigma}  \mathcal{\hat{P}}_f \Big( \hat{f}_{i \mu \sigma}^{\dagger} \hat{b}_j^{\dagger} \hat{b}_i \hat{f}_{j \mu \sigma} + h.c. \Big)  \mathcal{\hat{P}}_f
    + \Big(J_{\perp} - \frac{3}{2} \frac{t_{\parallel}^2}{J_{\perp}} \Big) \sum_{j \mu} \hat{n}_{j\mu}^f 
    + \frac{3}{2} \frac{t_{\parallel}^2}{J_{\perp}} \sum_{\langle i,j \rangle} \sum_{\mu \mu'} \hat{n}_{i\mu}^f \hat{n}_{j\mu'}^f \notag\\
    &\quad- 4 \frac{t^2_{\parallel}}{V} \sum_{\langle i,j \rangle} \Big( \boldsymbol{-} \mathbf{\hat{J}}_i \cdot \mathbf{\hat{J}}_j + \frac{1}{4} \Big) \Big( \mathbf{\hat{S}}_i \cdot \mathbf{\hat{S}}_j + \frac{3}{4} \hat{n}_i^f \hat{n}_j^f \Big) + \big( V+J_{\perp} \big) \sum_{i} \hat{b}_i^{\dagger} \hat{b}_i \notag\\
    &\quad + \mu_h \Big( \sum_{j} \sum_{\mu \sigma} \hat{n}_{j\mu\sigma}^f +
    2\sum_{j} \hat{n}_j^b - N_h \Big)
\end{align}
\noindent
with the last term, we have introduced the chemical potential $\mu_h$, which controls the number of holes $N_h$ in the system.
\noindent Now, to derive an analytical expression for $\langle \hat{H} \rangle$, we consider the Hamiltonian in momentum space by performing the Fourier transform $ \hat{f}_{j\mu\sigma}^{\dagger} = \frac{1}{\sqrt{L}} \sum_k \hat{f}^{\dagger}_{k\mu\sigma} e^{-ikj} $ and 
 $ \hat{b}_{j}^{\dagger} = \frac{1}{\sqrt{L}} \sum_k \hat{b}^{\dagger}_{k} e^{-ikj} $
 where we have a 1-dimensional system of size \textit{L}. This gives us
\begin{align} \label{Appendix:H in k}
    \hat{\mathcal{H}}_{\text{eff},k}^{sc,cc} &= 
    - \frac{t_{\parallel}}{\sqrt{L}} \sqrt{2} \sum_{k q} \sum_{\mu \sigma} (-1)^{\sigma} \big(\hat{f}^{\dagger}_{k \mu \sigma}  \hat{f}^{\dagger}_{q-k \thickbar{\mu} \thickbar{\sigma}} \hat{b}_q \, \cos(k) + \text{h.c.} \big) + t_{\parallel} \sum_{k} \sum_{\mu \sigma} \cos(k) \hat{f}^{\dagger}_{k \mu \sigma} \hat{f}_{k \mu \sigma} \notag\\
    &\quad - 2\frac{t_{\parallel}}{L} \sum_{k k' q} \sum_{\mu \sigma} \cos(q) \hat{f}_{k \mu \sigma}^{\dagger} \hat{b}_{k'}^{\dagger} \hat{b}_{k+q} \hat{f}_{k'-q \mu \sigma} 
    + \Big(J_{\perp} - \frac{3}{2} \frac{t_{\parallel}^2}{J_{\perp}} \Big) \sum_{k \mu} \hat{n}_{k\mu}^f \notag\\
    &\quad + \frac{3}{2} \frac{t_{\parallel}^2}{J_{\perp} L} \sum_{kk'q} \sum_{\substack{\mu \mu' \\ \sigma \sigma'}} \hat{f}^{\dagger}_{k\mu\sigma} \hat{f}^{\dagger}_{k'\mu'\sigma'} \hat{f}_{k'-q \mu'\sigma'} \hat{f}_{k+q \mu\sigma} \, e^{-iq}
    \notag\\
    &\quad - \frac{t^2_{\parallel}}{V L} \sum_{kk'q} \sum_{\substack{\mu \mu' \\ \sigma \sigma'}} \hat{f}^{\dagger}_{k\mu\sigma} \hat{f}^{\dagger}_{k'\thickbar{\mu}\thickbar{\sigma}} \hat{f}_{k'-q \mu'\sigma'} \hat{f}_{k+q \thickbar{\mu}'\thickbar{\sigma}'} \, e^{-iq}
    +  \big(V+J_{\perp} \big) \sum_{k} \hat{b}_k^{\dagger} \hat{b}_k \notag \\
    &\quad + \mu_h \Big( \sum_{k} \sum_{\mu \sigma} \hat{f}^{\dagger}_{k \mu \sigma} \hat{f}_{k \mu \sigma} +
    2\sum_{k} \hat{b}_k^{\dagger} \hat{b}_k - N_h \Big)
\end{align}
\noindent
where we have neglected the Gutwiller projections, which is an acceptable approximation in the low-doping regime $\delta < 0.5$ where the overlap between different spinon-chargons singlets is negligible.
\section{Derivation of the ground state parameters} \label{appendix: gs parameters}
\noindent The mean-field theory that we provide is based upon an ansatz wavefunction which captures the essential nature of the crossover: the tensor product between the BCS wavefunction where the Cooper pairs are formed by two spinon-chargons on opposite layers in a spin singlet state and the coherent state of chargon-chargons at zero quasi-momentum

\begin{equation} \label{Appendix:ansatz wavefunction}
    \ket{\Psi} = \bigg[ \underbrace{\prod_k \Big(u_k + v_k \frac{1}{\sqrt{2}} \sum_{\sigma} (-1)^{\sigma} \hat{f}^{\dagger}_{k1\sigma} \hat{f}^{\dagger}_{-k0\thickbar{\sigma}} \Big)}_\text{sc BCS} \otimes \underbrace{\mathcal{N} \exp\big({\beta \hat{b}_{k=0}^{\dagger}} \big)}_\text{cc BEC} \bigg] \ket{0}
\end{equation}
\noindent where the first term represents the BCS nature of spinon-chargons singlets with $u_k, v_k \in \mathbb{C}$ parameters. The second part describes the BEC-like chargon-chargon regime with normalization constant $\mathcal{N}$ and order parameter $\beta \in \mathbb{C}$. The normalization requirement for the BCS term yields 
\begin{equation} \label{Appendix:Normalization}
    \prod_k \big( |u_k|^2 + |v_k|^2 \big) \overset{!}{=} 1
\end{equation}
\noindent
which is certainly satisfied if we demand $|u_k|^2 + |v_k|^2 = 1$ $\forall \, k$.
The ground-state values of the parameters can be determined variationally by minimizing the expected value of the energy $\nabla \langle \Psi|\hat{H}|\Psi \rangle \overset{!}{=} 0$. Hence, firstly, we compute the contribution to $\langle \hat{H} \rangle$ for each term in the Hamiltonian. We give an example: consider the second term in Eq.(\ref{Appendix:H in k}). This term acts trivially on the BEC part of the wavefunction, and so we focus on the following expression
\begin{equation}
   \bra{0} \prod_p \Big(u_p^{\ast} + v_p^{\ast} \frac{1}{\sqrt{2}} \sum_{s} (-1)^{s} \hat{f}_{-p0\thickbar{s}'} \hat{f}_{p1s} \Big) \,
    \hat{f}^{\dagger}_{k \mu \sigma} \hat{f}_{k \mu \sigma} \,
    \prod_{p'} \Big(u_{p'} + v_{p'}  
    \frac{1}{\sqrt{2}} \sum_{s'} (-1)^{s'} \hat{f}^{\dagger}_{p'1s'} \hat{f}^{\dagger}_{-p'0\thickbar{s}'} \Big) \ket{0}
\end{equation}
\noindent
A special role is played by the terms $p=p'=k$ for which we have
\begin{equation}
    \bra{0} \Big(u_k^{\ast} + v_k^{\ast} \frac{1}{\sqrt{2}} \sum_{s} (-1)^{s} \hat{f}_{-k 0 \thickbar{s}} \hat{f}_{k 1 s} \Big) \,
    \hat{f}^{\dagger}_{k \mu \sigma} \hat{f}_{k \mu \sigma} \,
    \Big(u_k + v_k  
    \frac{1}{\sqrt{2}} \sum_{s'} (-1)^{s'} \hat{f}^{\dagger}_{k 1 s'} \hat{f}^{\dagger}_{-k 0 \thickbar{s}'} \Big) \ket{0}
\end{equation}
\noindent
where the terms contributing are the ones that ensure that the spinon-chargons are first created and then annihilated.
If we consider the case $\mu = 1$ and $\sigma = \; \uparrow $ the only non-zero contribution is
\begin{equation}
   \bra{0} 
   \wick[offset=1.2em,sep=0.4em]{
   \frac{1}{2} |v_k|^2 \c1{\hat{f}}_{-k0\downarrow}  \c2{\hat{f}}_{k 1 \uparrow} \,
    \c2{\hat{f}^{\dagger}}_{k 1 \uparrow} \c2{\hat{f}}_{k 1 \uparrow} \,
    \c2{\hat{f}^{\dagger}}_{k 1 \uparrow} \c1{\hat{f}}^{\dagger}_{-k 0 \downarrow}} \ket{0}
\end{equation}
\noindent 
The same applies for $\mu = 1$ and $\sigma = \; \downarrow $. The procedure is analogous in the case of $\mu = 0$, but now, we must consider the terms $p=p'=-k$. Hence, taking into account the four possible combinations of $\mu$ and $\sigma$, we obtain the total contribution for the second term in the Hamiltonian
\begin{equation}
    \bra{\Psi} t_{\parallel} \sum_{k \mu \sigma} \cos(k) \hat{f}^{\dagger}_{k \mu \sigma} \hat{f}_{k \mu \sigma} \ket{\Psi} =  t_{\parallel} \sum_k \big(|v_k|^2 + |v_{-k}|^2 \big)  \cos(k)
\end{equation}
\noindent The same procedure can be used to determine the contribution from the other terms in the Hamiltonian from Eq.(\ref{Appendix:H in k}). The total expectation value for the energy is 
\begin{align} \label{Appendix:full energy}
    \langle \hat{H} \rangle &=  
    -2\frac{t_{\parallel}}{\sqrt{L}} \sum_k \big( v_k^{\ast}u_k \beta \cos(k) + v_{-k}^{\ast}u_{-k} \beta \cos(k) + \text{c.c} \big) \notag\\
    & \quad + \big( t_{\parallel} -2\frac{t_{\parallel}}{L} |\beta|^2 \big) \sum_k \big(|v_k|^2 + |v_{-k}|^2 \big)  \cos(k) 
    + \Big(J_{\perp} - \frac{3}{2} \frac{t_{\parallel}^2}{J_{\perp}} + \mu_h \Big) \sum_k \big(|v_k|^2 + |v_{-k}|^2 \big) \notag\\
    & \quad + \frac{3}{2} \frac{t_{\parallel}^2}{J_{\perp} L} \sum_{kk'} \Big[2 \cos(k-k') (v_k^{\ast} u_k u_{k'}^{\ast} v_{k'}) \, + \notag \\
    & \qquad \qquad \qquad \qquad + |v_k|^2 |v_{k'}|^2 + |v_{-k}|^2 |v_{-k'}|^2 + |v_{-k}|^2 |v_{k'}|^2 + |v_k|^2 |v_{-k'}|^2 \Big] \notag\\
    & \quad + \big(V+J_{\perp} + 2\mu_h \big) |\beta|^2 - \mu_h N_h
\end{align}
So far, we haven't made any assumptions on $v_k$ and $u_k$ except for the normalization requirement. However, since $ \langle \hat{H} \rangle$ has to be real, we can choose w.l.o.g. $v_k, u_k \in \mathbb{R}$, which we will assume from now on. The terms of fourth order in $v_k$ in Eq.(\ref{Appendix:full energy}) correspond to the lowest order mean-field Hartree terms, which can be neglected since they
vanish in the limit of short-range interactions \cite{Parish_2014}. Then Eq.(\ref{Appendix:full energy}) simplifies to
\begin{align}
    \langle \hat{H} \rangle &=  
    - 4 \frac{t_{\parallel}}{\sqrt{L}} \sum_k \big( v_k u_k \beta \cos(k) + \text{c.c} \big) \notag\\
    & \quad + 2 \big( t_{\parallel} - 2 \frac{t_{\parallel}}{L} |\beta|^2 \big) \sum_k\cos(k) |v_k|^2 
    + 2 \Big(J_{\perp} - \frac{3}{2} \frac{t_{\parallel}^2}{J_{\perp}} + \mu_h \Big) \sum_k |v_k|^2 \notag\\
    & \quad +  3 \frac{t_{\parallel}^2}{J_{\perp} L} \sum_{kk'} \cos(k-k') (v_k u_k u_{k'} v_{k'}) + \big(V+J_{\perp} + 2\mu_h \big) |\beta|^2 -\mu_h N_h
\end{align}
In order to determine the ground state properties of the system, we need to minimize $\langle \hat{H} \rangle$. We can exploit the normalization restriction from Eq. (\ref{Appendix:Normalization}) to parametrize the BCS coefficients as $v_k = \sin\tfrac{\theta_k}{2}$ and $u_k = \cos \tfrac{\theta_k}{2}$, then
\begin{align}
    \langle \hat{H} \rangle &= 
    -2 \frac{t_{\parallel}}{\sqrt{L}} \sum_k \Big[ \sin\theta_k \; \beta \cos(k) + \text{c.c} \Big]
    + \Big(2 t_{\parallel} - 4 \frac{t_{\parallel}}{L} |\beta|^2 \Big) \sum_k \sin^2\frac{\theta_k}{2}\, \cos(k) \notag\\
    & \quad + 2 \Big(J_{\perp} - \frac{3}{2} \frac{t_{\parallel}^2}{J_{\perp}} + \mu_h \Big) \sum_k \sin^2\frac{\theta_k}{2} \notag\\
    & \quad + \frac{3}{2} \frac{t_{\parallel}^2}{J_{\perp} L} \sum_{kk'} \frac{1}{2} \sin\theta_k \; \sin\theta_{k'} \cos(k-k') + \big(V+J_{\perp} + 2\mu_h \big) |\beta|^2 -\mu_h N_h
\end{align}
\subsubsection*{\textbf{BCS ground state parameter}}
\noindent
As a first step, we want to find the stationary point with respect to the BCS coefficient $\theta_k$ 
\begin{align}  \label{minimizing by theta}
    \pdv{\langle \hat{H} \rangle}{\theta_k} &= 
    - 2 \frac{t_{\parallel} }{\sqrt{L}} \Big[ \cos \theta_k \;\beta  \cos(k) + \text{c.c} \Big] 
    + \Big(t_{\parallel} - \frac{t_{\parallel}}{L} |\beta|^2 \Big) \sin \theta_k \; \cos(k) \notag\\
    &\quad + \Big(J_{\perp} - \frac{3}{2} \frac{t_{\parallel}^2}{J_{\perp}} + \mu_h \Big) \sin \theta_k \notag\\
    &\quad + \frac{3}{2} \frac{t_{\parallel}^2}{J_{\perp} L}  \sum_{k'} \cos \theta_k \; \sin \theta_{k'} \cos(k-k') \overset{!}{=} 0
\end{align}
\noindent
For simplicity, we define the quantities 
\begin{align}
    \xi_k &:= \Big( t_{\parallel} - \frac{t_{\parallel}}{L} |\beta|^2 \Big) \cos(k) + \Big(J_{\perp} - \frac{3}{2} \frac{t_{\parallel}^2}{J_{\perp}} + \mu_h \Big) \notag\\
    V_{kk'} &:=  \frac{3}{2} \frac{t_{\parallel}^2}{J_{\perp}} \cos(k-k') 
\end{align}
\noindent
and parametrize $\theta_k$ by 
\begin{equation} \label{Appendix:Eq Delta}
  2 v_k u_k =  \sin \theta_k = \frac{\Delta_k}{\sqrt{\xi_k^2 + \Delta_k^2}}, \qquad u_k^2 - v_k^2 = \cos \theta_k = \frac{\xi_k}{\sqrt{\xi_k^2 + \Delta_k^2}}
\end{equation}
\noindent
From this parametrization, it is straightforward to show that 
\begin{align}
    v_k^2 &= \frac{1}{2}\Big( 1 -  \frac{\xi_k}{\sqrt{\xi_k^2 + \Delta_k^2}} \Big) \notag\\
    u_k^2 &= \frac{1}{2} \Big( 1 + \frac{\xi_k}{\sqrt{\xi_k^2 + \Delta_k^2}} \Big)
\end{align}
\noindent
Substituting into Eq.(\ref{minimizing by theta}) yields
\begin{equation}
     0 = - 2 \frac{t_{\parallel} }{\sqrt{L}} \Big[\beta \cos(k) + \text{c.c} \Big] +\Delta_k + \frac{1}{L} \sum_{k'} V_{k,k'} \frac{\Delta_{k'}}{\sqrt{\xi_{k'}^2 + \Delta_{k'}^2}} 
\end{equation}
\noindent
Finally, we can explicitly write down the \textit{Gap equation} for the BCS state of spinon-chargons
\begin{equation}
    \Delta_k = 2 \frac{t_{\parallel} }{\sqrt{L}} \Big[\beta \cos(k) + \text{c.c} \Big] -  \frac{1}{L} \sum_{k'} V_{k,k'} \frac{\Delta_{k'}}{\sqrt{\xi_{k'}^2 + \Delta_{k'}^2}} 
\end{equation}
\subsubsection*{\textbf{BEC ground state parameter}}
We now compute the stationary point with respect to the BEC parameter $\beta$
\begin{equation} \label{BEC minimization}
     \pdv{\langle \hat{H} \rangle}{\beta} = 
     - 4 \frac{t_{\parallel}}{\sqrt{L}} \sum_k  v_k u_k \cos(k)
     - 4 \frac{t_{\parallel}}{L} \beta^{\ast} \sum_k \cos(k) |v_k|^2 + \big(V+J_{\perp} + 2\mu_h \big)  \beta^{\ast} \overset{!}{=} 0
\end{equation}
\noindent
By previous considerations $v_k, u_k \in \mathbb{R}$ and $\beta \in \mathbb{C}$. Then, Eq.(\ref{BEC minimization}) can be split by separately considering its imaginary and real parts, which give us, respectively
\begin{align}
    \Re(\beta) &= \frac{2 \frac{t_{\parallel}}{\sqrt{L}}  \sum_k \frac{\Delta_k}{\sqrt{\xi_k^2 + \Delta_k^2}} \cos(k) }{\big(V + J_{\perp} + 2\mu_h \big) - 4 \frac{t_{\parallel}}{L} \sum_k \frac{1}{2}\Big( 1 -  \frac{\xi_k}{\sqrt{\xi_k^2 + \Delta_k^2}} \Big)\cos(k) } \notag\\
    \Im(\beta) &= 0
\end{align}
\noindent
where we have expressed the equations in terms of the BCS parameter $\Delta_k$.
\subsubsection*{\textbf{Holes number}}
In Eq. (\ref{Appendix:HAmiltonian real space}) we have introduced the chemical potential $\mu_h$. This allows us to fix the number of holes in the system by minimizing with respect to $\mu_h$
\begin{equation}
    \pdv{\langle \hat{H} \rangle}{\mu_h} = 2 \sum_k |v_k|^2 + 2|\beta|^2 - N_h \overset{!}{=} 0
\end{equation}
where $N_h$ is the number of holes in the system. Thus, we have an additional equation relating the BEC parameter $\beta$ and the BCS parameter $\Delta_k$. Explicitly, we have
\begin{equation}
    N_h =  2 \sum_k \frac{1}{2}\Big( 1 -  \frac{\xi_k}{\sqrt{\xi_k^2 + \Delta_k^2}} \Big) + 2|\beta|^2
\end{equation}
\subsubsection*{\textbf{Observables}}
\noindent The energy with respect to the ansatz ground state wave function in Eq. (\ref{Eq:ansatz wavefunction}) can be expressed in terms of the new parameter $\Delta_k$
\begin{align}
    \langle \hat{H} \rangle &= -2 \frac{t_{||}}{\sqrt{L}} \sum_k \Big[ \frac{\Delta_k}{\sqrt{\xi_k^2 + \Delta_k^2}} \beta \cos(k) + c.c \Big] + 2 (t_{||} - 2\frac{t_{||}^2}{L}|\beta|^2) \sum_k \frac{1}{2}\Big( 1 -  \frac{\xi_k}{\sqrt{\xi_k^2 + \Delta_k^2}} \Big) \cos(k) \notag \\
    &+ 2 \Big(J_{\perp} - \frac{3}{2} \frac{t_{\parallel}^2}{J_{\perp}} + \mu_h \Big) \sum_k  \frac{1}{2}\Big( 1 -  \frac{\xi_k}{\sqrt{\xi_k^2 + \Delta_k^2}} \Big) + \frac{3}{4} \frac{t_{||}^2}{J_{\perp} L} \sum_{k k'} \frac{\Delta_k \Delta_{k'}}{\sqrt{\xi_k^2 + \Delta_k^2} \sqrt{\xi_{k'}^2 + \Delta_{k'}^2}} \cos(k-k') \notag \\
    &+ (V + J_{\perp} +2\mu_h) |\beta|^2 - \mu_h N_h
\end{align}
\noindent Moreover, the sc and cc densities are straightforward to obtain $n_f = \frac{2}{L} \sum_k \frac{1}{2}\Big( 1 -  \frac{\xi_k}{\sqrt{\xi_k^2 + \Delta_k^2}} \Big)$ and $n_b = \frac{|\beta|^2}{L}$.
\section{Binding energy} \label{Appendix: binding E}
\noindent For the bilayer model under consideration, the binding energy is defined as
\begin{equation}\label{Appendix:binding energy}
    E_B(N_h) = 2(E_{N_h-2} - E_{N_h-1}) - (E_{N_h-2}-E_{N_h})
\end{equation}
where $N_h$ is the number of holes in the system, added in an alternating fashion to each layer.
In order to compute the binding energy, we consider the BEC limit of the crossover where the holes in the system tend to form tightly bound pairs of chargons. Here, the cc's are described by a non-interacting BEC which is characterized by a vanishing chemical potential $\mu_{\rm BEC} \simeq 0$. Therefore, the addition to the system of a new bosonic pair (two holes) leaves the ground state energy approximately unchanged: $E_{N_h-2}-E_{N_h} \approx 0$. If only one extra hole is introduced to the system, then it will contribute to the formation of a spinon-chargon with an energy change given by $E_{N_h-1}-E_{N_h} = \Delta_{\rm gap}$ where $\Delta_{\rm gap}$ is the pairing gap emerging from the formation of sc's Cooper pairs.\\
The binding energy obtained from Eq.(\ref{Appendix:binding energy}) is thus $E_b = 2\Delta_{\rm gap}$. Here, we briefly discuss how to determine the value of the pairing gap starting from the ansatz wavefunction in Eq.(\ref{Appendix:ansatz wavefunction}) and focusing on the BCS term $|\Psi \rangle_{\rm BCS}=\prod_k \Big(u_k + v_k \frac{1}{\sqrt{2}} \sum_{\sigma} (-1)^{\sigma} \hat{f}^{\dagger}_{k1\sigma} \hat{f}^{\dagger}_{-k0\thickbar{\sigma}} \Big)\ket{0}$ representing a Gaussian state, which generally is the ground state of a quadratic Hamiltonian. Such Hamiltonian can be expressed using the Nambu spinor formalism
\begin{equation}
    \mathcal{\hat{H}} = \sum_{k} \hat{\psi}_k^{\dagger} ( \vec{h}_{k} \cdot \vec{\sigma})\hat{\psi}_{k} 
\end{equation}
with $\hat{\psi}_k$ the Nambu spinor, $\vec{\sigma} = (\sigma_x, \sigma_y, \sigma_z)$ the Pauli matrices and $\vec{h}_k = (\Delta_{k}^{(1)}, \Delta_{k}^{(2)}, \xi_k)$ where $\Delta_k = \Delta_{k}^{(1)}-i\Delta_{k}^{(2)}$. Here, the Hamiltonian can be easily diagonalized and the eigenvalues are $\pm E_k = \pm \sqrt{\Delta_k^2+\xi_k^2}$. Therefore, the energy gap is given by $\Delta_{\rm gap} = \min_k(E_k)$.

\section{Pairing function for 2D bilayer systems}
\begin{figure}
    \centering
    \includegraphics[width=0.6\linewidth]{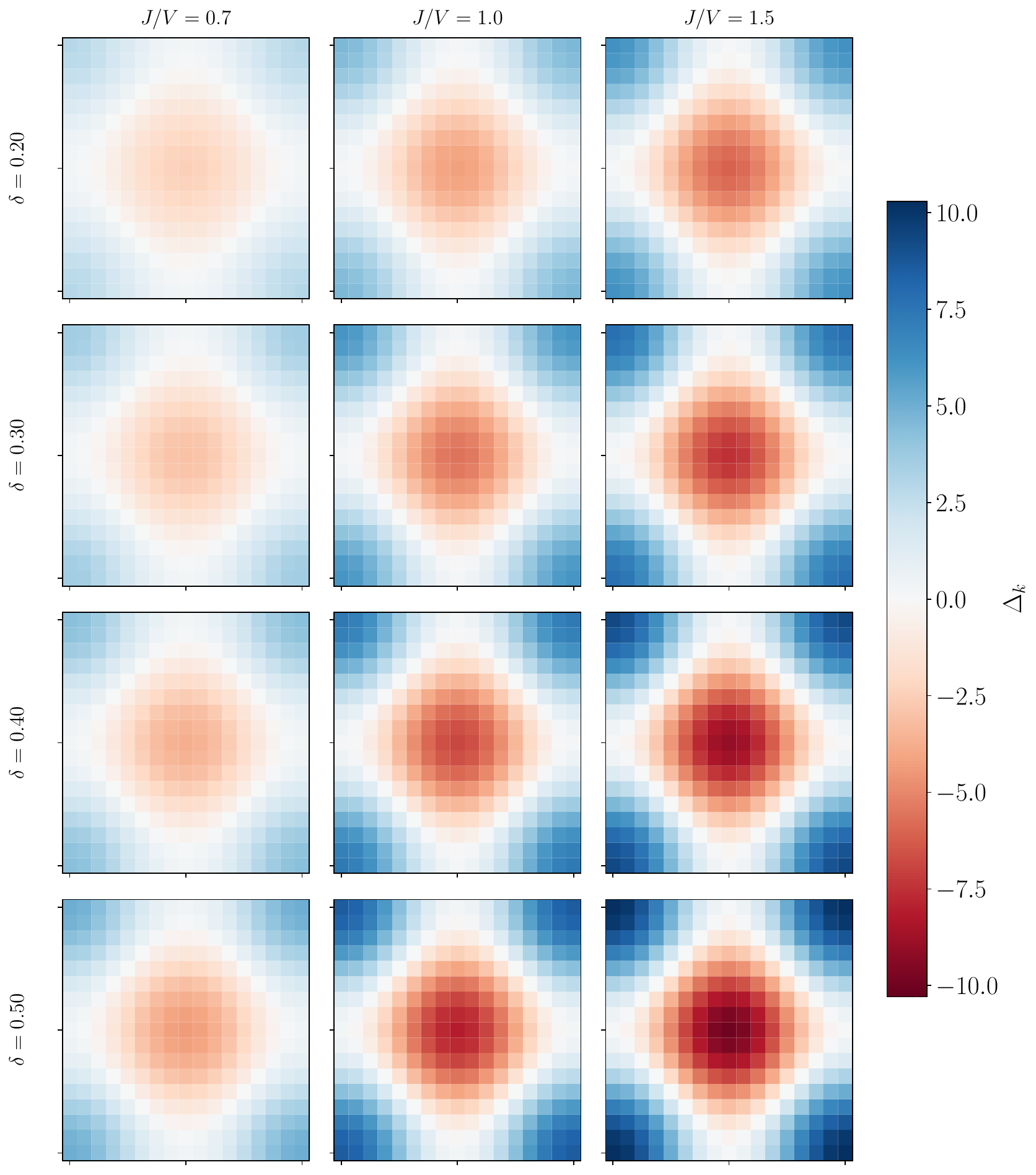}
    \caption{We show the pairing function $\Delta_k$ for the BCS ansatz wavefunction in a 2D bilayer system, as defined in the main text. Here we consider a system of size $L_x = L_y = 16$ for doping $\delta = \{\frac{52}{256},\frac{76}{256}, \frac{89}{256}, \frac{128}{256} \}$ at fixed $t_\parallel=0.1V$ at $J_\perp = \{0.7,1,1.5\}V$.}
    \label{fig:multiple_pairing_function_2D}
\end{figure}

\noindent
The mean-field analysis described Appendix \ref{appendix: gs parameters} can be directly extended to the 2D setting. Such extension is supported by the comparison with the DMRG simulation in the quasi-2D setting with a two-ladder system as shown in (reference the comparison). In Fig. \ref{fig:multiple_pairing_function_2D} we show the BCS pairing function \(\Delta_k\) for the 2D MixD \(t\!-\!J\) model on a \(16 \times 16\) square lattice with \(t_\parallel = 0.1V\), for various values of the ratio \(\frac{J_\perp}{V}\) and at multiple doping levels. The results show that the system consistently exhibits an extended \(s\)-wave pairing symmetry, with the magnitude of \(\Delta_k\) increasing as \(\frac{J_\perp}{V}\) grows. This trend is consistent with expectations for a BCS–BEC crossover, where the pairing gap becomes larger as the system transitions toward the BEC regime. 

\section{Details of the DMRG simulations} \label{Appendix: DMRG}
\noindent
The benchmarks in the main text are obtained using the single-site density matrix renormalization group (DMRG) algorithm implemented in the package SyTen \cite{syten1,syten2}. The implementation of the mixD model is based on the ones in Refs. \cite{Schloemer2023,lange2024pairingdomeemergentfeshbach}: We employ $U(1)_{N_{\mu=1}}\otimes U(1)_{N_{\mu=2}}\otimes U(1)_{S_z^\mathrm{tot}}$ (with $N_{\mu=i}$ the number of particles in chain $i$ and $S_z^\mathrm{tot}$ the total magnetization) in the DMRG ground state calculations. This corresponds to charge conservation in each individual leg (since $t_\perp=0$) and total magnetization conservation. As shown in the Appendix of Ref. \cite{Schloemer2023} this makes the ground state search much more efficient compared to calculations with only global charge conservation $U(1)_{N}\otimes  U(1)_{S_z^\mathrm{tot}}$. We use bond dimensions up to $\chi=1024$.\\
\end{document}